\documentclass[useAMS, usenatbib]{mn2e_letter}
\usepackage{graphicx}
\usepackage{microtype}
\usepackage{hyperref}

\usepackage{soul} 

\newcommand*{\hide}[1]{}

\newcommand{\mpc}{\rm M_\odot\;pc^{-2}}

\usepackage{amsmath, amsfonts, amssymb, fancyhdr}
\usepackage{natbib}

\newcommand{\cmc}{{\rm \; cm^{-3}}}
\newcommand{\kms}{{\rm \; km\;s^{-1}}}

\title{Bent Radio Jets Reveal a Stripped Interstellar Medium in NGC 1272}
\author[J. McBride and M. McCourt]{James McBride\thanks{E-mail: jmcbride@berkeley.edu, mkmcc@berkeley.edu} and Michael McCourt\\
Department of Astronomy, University of California, Berkeley, CA 94720-3411, USA}

\begin{document}
\maketitle
\begin{abstract}
We report the discovery of bent double jets in the elliptical galaxy NGC~1272, a member of the Perseus cluster. 
The jets have a radius of curvature of $\sim$2~kpc, much smaller than the galaxy half-light radius of $\sim$11~kpc.
This bending is likely a result of ram pressure, and indicates that the intracluster gas enters deep within the galaxy and that the interstellar medium of NGC~1272 has been significantly removed. 
X-ray observations of the surrounding intracluster medium allow us to constrain the pressure within the jet.
We find that the standard assumptions of equipartition often used in interpreting other jets underestimate the pressure in the jets by a factor $\sim$30.
\end{abstract}

\section{Introduction} \label{intro}
Interaction with the intracluster medium (ICM) in galaxy clusters likely depletes the interstellar media (ISM) of the cluster's member galaxies \citep[e.g.,][]{Dressler1984,Haynes1984}. 
This process plays an important role in the evolution of cluster galaxies, in part by helping quench star formation \citep[e.g.,][]{Boselli2006}. 
ICM/ISM interaction also affects several other observed properties of cluster galaxies.
For example, \citet{Chung2009} found evidence for ongoing ram pressure stripping of spiral galaxies in the Virgo cluster via long tails of H~{\sc i} and galaxies with H~{\sc i} disks smaller than their stellar disks.
\citet{Murphy2009} compared observed radio continuum emission with expectations from the far-infrared--radio correlation, and found six Virgo cluster members whose total radio emission was enhanced, but which had radio deficits along their outer edges. They interpreted their observations as evidence of interaction between the ICM and the galactic magnetic fields.

The jets launched by active galactic nuclei (AGN) in some cluster galaxies also demonstrate interactions between cluster galaxies and the ICM. 
\citet{Ryle1968} discovered the first such examples in two members of the Perseus cluster, IC~310 and NGC~1265. 
Each galaxy features an extended ($\sim$0.1--1~Mpc) tail on one side of the galaxy.
These are now known as ``narrow-angle tail galaxies'' or as ``head-tail galaxies'', and have been found in many other galaxy clusters \citep{Sarazin1988}. 
Head-tail sources are usually interpreted as an AGN-driven jet swept back by ram pressure as it moves through the ICM \citep{Miley1972,Owen1976,Begelman1979,Morsony2013}. 
Some galaxies with less extreme bending, and which are often but not always associated with central galaxies in clusters, have been called wide-angle-tail galaxies \citep{Sarazin1988}. 
We focus on sources with radio morphologies that likely result from the motion of their galaxy through the ICM, regardless of the degree of bending; we refer to all such sources as ``bent-doubles''.
The association between bent-doubles and galaxy clusters is so strong that bent-doubles have been used as a tool for discovering galaxy clusters in radio surveys \citep{Blanton2001,Wing2011}, and to make estimates of gas densities in intragroup and intrafilament environments \citep{Freeland2008,Edwards2010}.

The majority of bent-doubles have radii of curvature $R \sim$~10--100~kpc, though single-tailed radio sources associated with galaxies in clusters may be spatially unresolved jets bent on scales $<$10~kpc \citep{ODea1985}. 
The model proposed by \citet{Begelman1979}, and others like it, have  generally not considered the presence of an ISM in the host galaxy. 
\citet{Jones1979}, however, argued that elliptical galaxies in clusters should be expected to retain a significant ISM, and that bent-double morphologies could be produced via a central jet, pressure gradients in the ISM, and ICM ram pressure. 
In either scenario, bending the jets requires that they feel the headwind of the ICM, indicating that the ISM of the host galaxy must be removed on scales comparable to the radius of curvature. 
If the single tailed sources of \citet{ODea1985} are truly unresolved bent-doubles with small bending radii, they may represent galaxies from which the ISM has been very effectively removed.

We report the serendipitous discovery of bent radio jets associated with the galaxy NGC~1272, a massive elliptical galaxy in the Perseus cluster. 
The jets of NGC~1272 are bent within $\sim$2~kpc of the galaxy centre, and their morphology is consistent with being distorted by ram pressure from the motion of NGC~1272 through the ICM.
The bent jets provide compelling evidence that the ICM influence extends deep within the potential of the galaxy, and 
highlight the strong interactions between cluster galaxies and their gaseous environment. 
Additionally, standard assumptions about equipartition imply a pressure within the jet that is much lower than the pressure of the surrounding medium. 
We interpret this to mean that the magnetic fields and relativistic particles are not in equipartition, or that the ratio of relativistic protons to electrons is much greater than 1.

\section{Data} \label{data}

\begin{table*}
  \centering
  \begin{tabular}{r r r l}
    Centre frequency & Bandwidth & Observing time & Archive file IDs \\
    (GHz)            & (GHz)     & (minutes)      & \\ \hline
    1.436            & 0.128     & 70             & \parbox{6.5cm}{L\_osro.55780.593262442126, \\ L\_osro.55798.593668483794,\\ L\_osro.55807.66087910879} \\[2em]
    1.804            & 0.128     & 70             & Same observations as 1.436 GHz \\[1em]
    3.148            & 0.256     & 95             & \parbox{6.5cm}{sysstartS\_001.56239.18234083333, \\ sysstartS\_003.56242.428367500004, \\ sysstartS\_000.56272.34511214121, \\ sysstartS\_000.56279.35494653935}  \\[3em]
    5.164            & 1.152     & 40             & \parbox{6.5cm}{C\_quad2\_000.55624.950303923615,\\ C\_quad4.55629.98669633102, \\ C\_quad4.55630.97231834491, \\ C\_quad1.55638.906346400465} \\
  \end{tabular}
  \caption{Summary of JVLA data from the NRAO archive that we present. 
  The L band frequencies (1.436~GHz and 1.804~GHz) were observed simultaneously, but are listed separately because they are non-contiguous, and were individually self calibrated and imaged.}
\label{tab:obs}
\end{table*}

The compact radio source 3C84 is located in the nucleus of NGC~1275, the central galaxy in the Perseus Cluster.\footnote{Throughout the paper, we will refer to the central radio source as 3C84, the central galaxy as NGC~1275, and the cluster itself as Perseus.} Because 3C84 is one of the brightest objects in the sky at radio frequencies, it has been observed as part of operations/maintenance of the Jansky Very Large Array (JVLA).
These data are all publicly available via the National Radio Astronomy Observatory Science Data Archive (\href{https://archive.nrao.edu/archive/advquery.jsp}{NRAO archive}); we present a subset of this data here.
We searched the NRAO archive for observations of 3C84 in four bands: 90~cm (P band), 20~cm (L band), 13~cm (S band), and 6~cm (C band). 
For P, L, and S band, we searched for data in the A configuration. It is the most extended JVLA configuration, and thus provides the highest possible resolution at any given frequency. 
For C band, we searched for data in the B configuration, which is roughly a factor of three more compact than the A configuration. This gives an angular resolution that is comparable to that of L and S band in the A configuration. 

We ultimately selected $\sim$1 hour of data at L band, $\sim$1.5 hours of S band data, and $\sim$0.5 hours of C band data. 
We provide a more detailed summary in Table \ref{tab:obs}.
There were no usable data on the NRAO archive at P band. 
In each band/configuration, we selected data from the NRAO archive on the basis of four primary factors: moderately long integration times on 3C84, a low number of missing antennae or automatically flagged data, frequency configurations consistent with already selected data, and observations taken within a $\sim$month of already selected data.

We used the standard AIPS tasks for data reduction and calibration.
The bright radio source 3C84 was the nominal target of all observations, and thus served as a natural phase calibrator. 
3C84 is moderately variable, and thus does not make an ideal flux calibrator; unfortunately, however, our data contain no observations of standard flux calibrators. 
Cross checking the fluxes we measured for other sources in the field-of-view with published fluxes indicates that the absolute flux scale is correct to within $\lesssim$10\%.
We factor this uncertainty into our estimates of the fluxes of the bent jets in NGC~1272.
We used self calibration of 3C84 to improve the dynamic range of the images at each frequency. 
However, limited $uv$-coverage in each band prevented us from achieving the current state-of-the-art dynamic ranges possible with the upgraded JVLA.

\section{Results \& Discussion}
\begin{figure*}
  \includegraphics[width=7.0in]{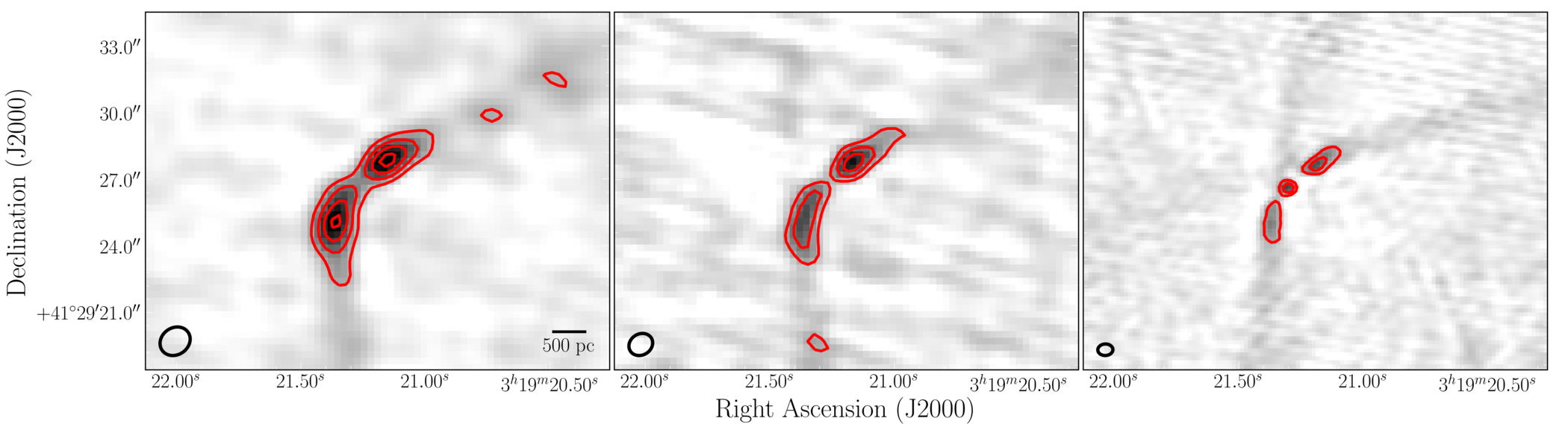}
  \caption{Contours (red) for each of the three frequencies where emission was detected: 1.44 GHz (left panel), 1.80 GHz (middle panel), 3.15 GHz (right panel). 
Contours are at 3, 6, 9, and 12 times the rms error. Grayscale shows the same data, but scaled such that whites represents -1 times the rms error, and black represents 9 times the rms error.
The FWHM of the beam at each frequency is in the lower left corner of each panel.
} \label{fig:three_lobes}
\end{figure*}
\subsection{Physical conditions within the jet}
We detect the jet in NGC~1272 at 1.44~GHz, 1.80~GHz, and 3.15~GHz. At 5.16~GHz, the jet was too faint to be detected with the available data. 
The residual errors in all of our maps are dominated by imperfect subtraction of sidelobes of 3C84, rather than thermal noise.
The images produced at each of the detected frequencies show clear evidence for bent radio jets, co-located with the centre of NGC~1272 to within the positional accuracy of the JVLA and existing optical/near-infrared observations of NGC~1272 ($\lesssim$0.3''). 
In the 3.15~GHz image, a core is distinctly visible, whereas the central source is blended with the jets in the 1.44~GHz and 1.80~GHz images (see Figure \ref{fig:three_lobes}). 
The gap between the core and the jets seen at 3.15~GHz is $\sim$500~pc; the absence of the gap at 1.4~GHz is likely due to the lower resolution at that frequency.
To determine the core flux at each frequency, we use the location of the core as seen at 3.15~GHz, and then sum the flux within the beam size centred on that location.
Remaining flux north and south of the core are included in the jet flux estimates.
Table \ref{tab:flux} provides the resulting flux measurements at each frequency. 

The morphology and position of the jets is consistent in all three bands, and is insensitive to uncertainty in the absolute flux calibration. 
Since the residual errors from calibration have different structure in each frequency band, the jet shape is unlikely to be an artifact of the imaging process.
The northern and southern jets have the same brightness at each frequency, within statistical uncertainty, so we expect the jets are either non-relativistic or are closely aligned with the plane of the sky.
The relative flux and bending of the two jets are consistent with motion through a surrounding medium \citep[e.g.,][]{Miley1972}. 
Radio jets that are bent by interaction with a dense, inhomogeneous medium in their galaxy are generally asymmetric and have factor $\sim$few differences in the flux on each side \citep[e.g.,][]{Dallacasa2013}.
\begin{table}
  \centering
  \begin{tabular}{r r r r}
    Band     & Core              & Northern jet     & Southern jet \\ \hline
    1.44 GHz & 1.0 $\pm$ 0.2 mJy & 4.0 $\pm$ 0.4 mJy & 4.7 $\pm$ 0.5 mJy \\
    1.80 GHz & 0.7 $\pm$ 0.2 mJy & 3.5 $\pm$ 0.5 mJy & 3.7 $\pm$ 0.6 mJy \\
    3.15 GHz & 0.4 $\pm$ 0.1 mJy & 1.2 $\pm$ 0.2 mJy & 0.9 $\pm$ 0.2 mJy \\
    5.16 GHz & 0.6 $\pm$ 0.4 mJy & 0.6 $\pm$ 1.5 mJy & 2.5 $\pm$ 1.7 mJy \\
  \end{tabular} \caption{Summary of measured fluxes.} \label{tab:flux}
\end{table}

Without accounting for projection effects, we estimate the physical extent of the jets. 
The jet length is $\ell$~$\sim 1.5$~kpc for both the northern and southern sides.
We fit a circle by eye to the shape of the emission, and estimate a radius of curvature of the jets $R \sim2$~kpc. 
The jet width $h$ is not clearly resolved at the frequencies we imaged.\footnote{A physical size $\sim$220~pc could be resolved at 5~GHz with A configuration data from the JVLA. 
Unfortunately, 3C84 has a relatively flat spectrum, whereas the flux from NGC~1272 is falling off faster than $\nu^{-1}$ above 3~GHz. 
We estimate a dynamic range 200,000:1 would be required to detect the jets of NGC~1272, which would be challenging to achieve with available archival data.}
Nonetheless, the observed jet width $h_{\rm obs}$ is larger than the Gaussian beam width, $b$, at each frequency. 
We thus estimate the true jet width via $h \sim (h_{\rm obs}^2 - b^2)^{1/2} \sim 220$~pc.
The maps at each frequency yield the same estimate of $h$ to within 10\%.

The flux in the core at each frequency is consistent with a power law $S_\nu \propto \nu^{-0.8}$, typical of astrophysical synchrotron emission. 
The fluxes in the jets at 1.44~GHz and 1.80~GHz follow the same power law slope, but the flux at 3.15~GHz falls below this power law; this falloff is consistent with a cooling break in the synchrotron emission. 
We use the 1.44~GHz and 1.80~GHz fluxes and the jet size to estimate the pressure and magnetic field in the jets, with the standard assumption of roughly equal energy density in the magnetic field and in relativistic particles \citep[typically called equipartition; see][]{Pacholczyk1970,Longair1994}. 
This assumption yields an expression for the pressure 
\begin{equation}
  P_{\rm min} = \left(\frac{7}{12}\right) 
  \left[\frac{ C_{12} L_{\rm rad} (1 + k)}{ (2 \pi)^{3/4} \phi V}\right]^{4/7}\; {\rm dynes\;cm}^{-2}.
\end{equation}
Here, $C_{12}$ is a constant that depends on the synchrotron spectral index and frequency cutoffs \citep{Pacholczyk1970}, $L_{\rm rad}$ is the radio luminosity, $k$ is the ratio of relativistic proton to relativistic electron energy, and $\phi V$ is the volume of the synchrotron emitting region. 
In interpretations of jets, it is typical to assume $k = 1$, a spectral index $\alpha \sim0.6$--1, and upper and lower frequency cutoffs of order 10~MHz and 10~GHz \citep[e.g.][]{Begelman1979,ODea1987a,Freeland2008}.
Using these assumptions and taking $\phi = 1$, we find $P_{\rm min} \sim 10^{-11}$ dynes cm$^{-2}$. 
The corresponding magnetic field strength in the jet for these assumptions is B~$\sim$~14~$\mu$G. 
If we interpret the sharp fall-off in flux between 1.80~GHz and 3.15~GHz as a synchrotron cooling break, the minimum age of the jets is $\sim$10~Myr.

This result is inconsistent with momentum conservation across the ICM/jet interface. 
The pressure in the jets should be comparable to the ram pressure, $\rho_{icm} v_{gal}^2$, ignoring terms of order $\mathcal{M}^{-2}$, where $\mathcal{M}$ is the Mach number of the galaxy through the ICM. 
The proton number density at the location of NGC~1272 within the cluster is $n_{icm} \sim 0.005 \cmc$ \citep{Sanders2004,Cavagnolo2009}. 
The line-of-sight velocity of NGC~1272 is $v_{los} \sim 1400 \kms$; we assume a total velocity $v_{gal} \sim 2000 \kms$. 
The ram pressure from the ICM acting on the jet is then $P_{\rm ram} \sim 3 \times 10^{-10}$ dynes cm$^{-2}$, a factor of $\sim$30 larger than $P_{\rm min}$. 
Under such a strong pressure imbalance, the jet should collapse on a timescale $\sim 10^5$~yr. 
One possible resolution is that the volume filling factor for synchrotron emission, $\phi$, could be significantly less than unity.
Pressure equilibrium $P_{\rm min} \sim P_{\rm ram}$ requires $\phi V$ to be $\sim$400 times smaller than we have assumed. 

Another possibility is that the relativistic protons dominate electrons, with $k >> 1$ (e.g., \citealt{Beck2005}; \citealt{Croston2014}, however, argue that proton dominated jets are unlikely).
If we require $P_{\rm min} = P_{\rm ram}$ and maintain the assumption that magnetic fields and relativistic particles have comparable energy densities, then we find $k \sim 800$. 
This assumption also yields $B_{\rm min} \sim 90$~$\mu$G, which corresponds to a cooling time $\sim$1~Myr. 

This result is consistent with previous analysis of equipartition assumptions in radio jets \citep{Hardcastle1998,Croston2003}, though to our knowledge this is the first such example in a bent-double.
This analysis shows that standard equipartition assumptions fail to explain the properties of the jets in NGC~1272. 
If this holds true more generally, it is significant for estimates of ram pressures, and hence gas densities, in intracluster, intragroup, and even intrafilament environments using observations of bent-doubles \citep[e.g.][]{Freeland2008,Edwards2010}. 
If we used $k = 1$ and other standard assumptions to interpret NGC~1272, we would estimate a gas density in Perseus more than an order of magnitude lower than the value found from x-ray observations \citep[e.g.,][]{Sanders2004}. 

\subsection{ISM/ICM interaction}
\begin{figure*}
  \centering
  \includegraphics[scale=0.16]{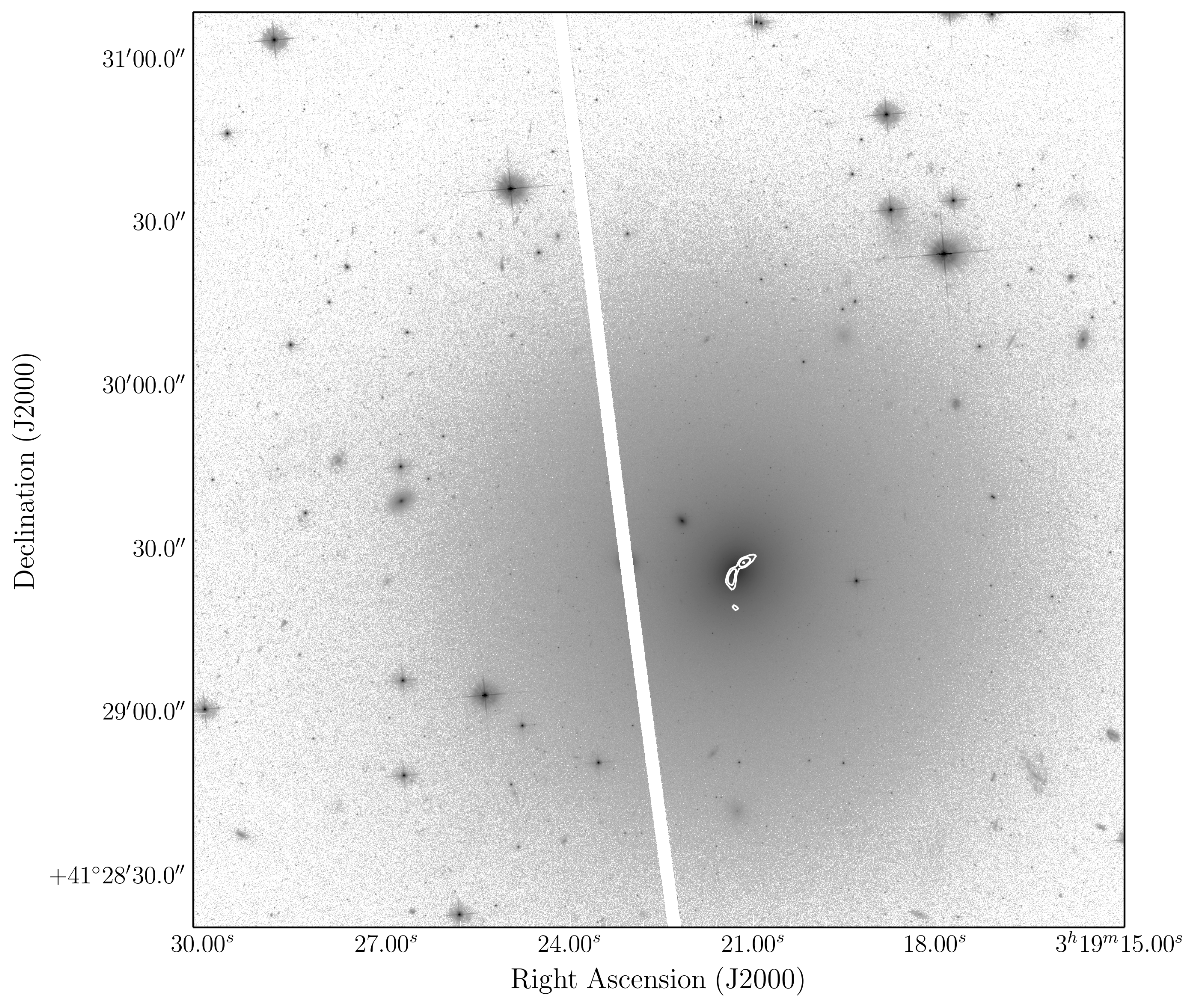}
  \caption{Radio contours from the combined 1.4~GHz and 1.8~GHz maps are shown in white, and are overlaid on a {\it Hubble Space Telescope} F555W image of NGC~1272. 
The white strip running through the middle of the image is a gap between the two plates in the Wide Field Camera.
Contour levels start at 0.5~mJy/beam, and increase by factors of two. 
The {\it HST} data are discussed in detail in \citet{deRijcke2009} and \citet{Penny2012}. 
} \label{fig:radio_overlay}
\end{figure*}

NGC~1272 is an elliptical galaxy, and is optically the second brightest galaxy in the Perseus cluster, following NGC~1275 \citep{deRijcke2009}. It has a projected separation from the cluster centre of 110~kpc and a cluster-centric line-of-sight velocity of $\sim$1400~$\kms$. Measurements of its half-light radius, $R_e$, range from 8.7--11.5~kpc \citep{Schombert1987,Hudson2001,Alonso2003}.
The radius of curvature of its jets, $R\sim$~2~kpc, is thus well within the galaxy, as shown in Figure \ref{fig:radio_overlay}; that the ICM can influence dynamics so deep within the galaxy is striking. 
In either the \citet{Begelman1979} model or \citet{Jones1979} model, the small bending radius of the jets requires NGC~1272 to have essentially no ISM material at radii of $\sim$2~kpc and beyond. 

The extreme degree of bending of the NGC~1272 jets may have some precedent in the literature.
\citet{ODea1985} compiled observations of 51 bent-doubles\footnote{They only included narrow-angle-tail galaxies; wide-angle-tail galaxies in non-cD galaxies were not included.} and found 12 sources that had a single-sided tail of radio emission. 
With the typical resolution of the observations, the single-tailed sources could be bent-doubles with bending radii $\lesssim$~5~kpc. 
\citet{ODea1985} discussed the possibility of one-sided jets powering the single-tailed sources, but their favored interpretation was that the single-tailed sources are unresolved bent-doubles from which the ISM has been stripped to small radii. 
The clearly resolved bent jets in NGC~1272 provide more conclusive evidence for ICM ram pressure acting at the very centre of a cluster galaxy. 

Due to its proximity to the cluster centre, NGC~1272 experiences stronger ram pressure than other known bent-doubles.
The Perseus cluster has five previously known bent-doubles with radii in the range 560--1480~kpc \citep{Sijbring1998}, compared to 110~kpc for NGC~1272.
Since the density falls roughly as $\rho \propto r^{-2}$ in the outskirts of the cluster \citep[e.g.,][]{Sarazin1988,Croston2008}, 
NGC~1272 experiences ram pressure at least an order of magnitude larger than any known bent-double in Perseus. 
Outside the Perseus cluster, \citet{ODea1987b} provide separations from cluster centre for the 51 bent-doubles in \citet{ODea1985}, including a few sources that have separations of $\lesssim$~100~kpc.
\hide{though \citet{ODea1987b} also noted uncertainties in cluster centre positions of 1'--2'.}
While the 3-d cluster-centric velocity of NGC~1272 is uncertain, it has a relative line-of-sight velocity that is well above the average for sources considered in \citet{ODea1985}.
The ram pressure acting on NGC~1272 plausibly exceeds that of any previously known bent-double.
Such strong ram pressure is consistent with NGC~1272 being among the most bent of known bent-doubles.

The total radio power in NGC~1272, from the jets and the nucleus, is also low relative to other bent-doubles, with $L_{1.44 {\rm GHz}}$~=~6.5~$\times 10^{21}$ W Hz$^{-1}$. 
This is comparable to one source in Perseus (T 74; \citealt{Sijbring1998}), but still smaller than any of the 51 bent-doubles compiled in \citet{ODea1985}.
The low luminosity of the bent-double in NGC~1272, and its proximity to the bright radio source 3C84, account for it escaping detection until now. 
The strong ram pressure experienced by NGC~1272 may play a role in its low luminosity.
Since powering the central AGN requires material to act as fuel, we consider the fate of remaining gas at smaller radii in NGC~1272.

\citet{Gunn1972} provided the original criterion for ram pressure removing interstellar gas from spiral galaxies in clusters. 
ISM is removed if
$  \rho_{icm} v_{gal}^2 > 2 \pi G \Sigma_* \Sigma_{\rm gas}$,
where $\Sigma_*$ is the stellar surface density, $\Sigma_{\rm gas}$ is the gas surface density, $\rho_{icm}$ is the ICM density, and $v_{gal}$ is the velocity of the galaxy relative to the ICM.
In an inhomogeneous medium, this expression holds separately for each phase of the ISM; molecular clouds, with higher surface densities, are more resilient to ram pressure stripping \citep[e.g.,][]{Fujita1999}.

\begin{figure}  \includegraphics[scale=0.080]{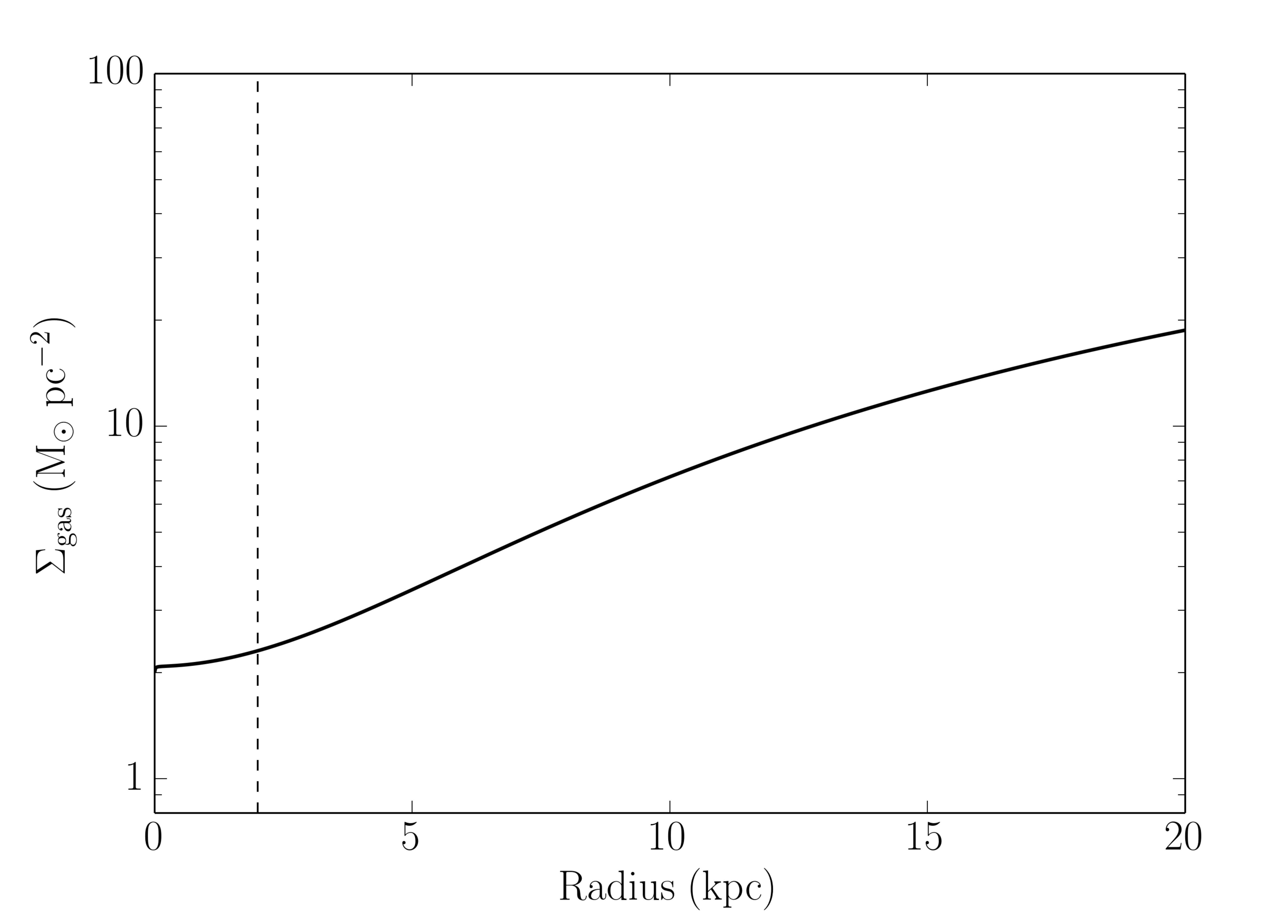}
  \caption{The ratio of gas mass to stellar mass for which ram pressure can strip the gas as a function of radius is represented by a solid black line. 
The dashed black line marks the radius of curvature of the jets, and thus the radius at which stripping has to have already taken place.
} \label{fig:sigmagas}
\end{figure}

To simplify the discussion, we consider a single gas phase with uniform $\Sigma_{\rm gas}$ within radius $r$.\footnote{Here, we are only considering gas bound to the galaxy, and not ICM gas co-located with the galaxy.}
Ram pressure stripping occurs at radii $r$ such that
\begin{equation}
    \rho_{icm} v_{gal}^2 \frac{r^2}{2 \pi G \Sigma_{\rm gas}(<r) M_{\rm tot}(<r)} \gtrsim 1. \label{eq:sigmagas}
\end{equation} 
As discussed above, we take measurements of $\rho_{icm} \sim 0.005 \cmc$ \citep{Sanders2004} and $v_{gal} \sim 2000 \kms$. 
We take mass profiles for the sum of the stars and dark matter, $M_{\rm tot}(<r)$, from published data and models. 
NGC~1272 has an apparent K$_s$ magnitude of 8.79 within a semimajor axis of 52'', or $\sim$19~kpc.
We use the K$_s$ magnitude to estimate the stellar mass within that radius to be 3~$\times 10^{11} M_\odot$ \citep{Bell2003}.
We take the \citet{Hudson2001} measurement of effective radius, $R_e \simeq$~11~kpc, as it was the median value we found. 
We assume that the stellar mass follows a Hernquist profile \citep{Hernquist1990} with this mass and effective radius.
Next, we estimate a dark matter mass using \citet{Moster2010}, and assume it follows an NFW density profile \citep{Navarro1997}. 
We then use Equation \ref{eq:sigmagas} to calculate the initial gas surface density such that the gas will be stripped down to a radius $r$.
Figure \ref{fig:sigmagas} plots this quantity as a function of $r$.
At 2~kpc, the radius of curvature of the jets, an initial gas surface density $\Sigma_{\rm gas} \lesssim 2\;\mpc$ is required for ram pressure to remove interstellar material. 
In the following paragraphs, we first consider whether this estimate of $\Sigma_{\rm gas}$ is consistent with current observations of NGC~1272. We then discuss whether the removal of gas in the past is consistent with observations of other elliptical galaxies. 

Direct information on the current gas content in NGC~1272 is limited.
\citet{Eskridge1991} placed an upper limit of $1.4 \times 10^{10} M_\odot$ on the H~{\sc i} mass in NGC~1272. 
To our knowledge, NGC~1272 has not been the target of molecular gas observations.
Active star formation would provide indirect evidence for the presence of gas, but constraints from star formation indicators like H$\alpha$, ultraviolet luminosity, and far-infrared luminosity are also limited. 
\citet{Moss2005} observed H$\alpha$ in galaxy cluster members, but did not detect emission from NGC~1272.
\citet{Surace2004} placed upper limits of 0.2 Jy, 0.4 Jy, 0.4 Jy, and 0.4 Jy for 12 $\mu$m, 25 $\mu$m, 50 $\mu$m, and 100 $\mu$m emission from NGC~1272, using {\em IRAS} data.
This corresponds to an upper limit on the star formation rate of 0.5~$M_\odot$~yr$^{-1}$.
NGC~1272 was detected by the Wide-field Infrared Survey Explorer, and has colors consistent with elliptical galaxies \citep{Jarrett2011}.
\citet{Marcum2001} used the Ultraviolet Imaging Telescope to observe a sample of bright galaxies, but did not detect 1500~\AA~emission from NGC~1272.
The Galaxy Evolution Explorer did detect 1500~\AA~and 2300~\AA~emission associated with the nucleus of NGC~1272, but the emission could be produced by the AGN, so does not provide a constraint on recent star formation. 
Thus we find no evidence in conflict with our limit on the current $\Sigma_{\rm gas}$ in NGC~1272. 

Observations of gas in galaxies similar to NGC~1272 support the plausibility of ram pressure stripping the ISM of NGC~1272 down to within 2~kpc of the galaxy centre. 
\citet{OcanaFlaquer2010} found a median molecular gas mass of 1.9~$\times 10^8 M_\odot$ in radio loud elliptical galaxies, and also observed a trend between $L_{CO}$ and $L_{FIR}$, as in typical star forming galaxies. 
If NGC~1272 had an initial gas mass equal to the median of their sample that was concentrated within 10~kpc, the corresponding surface density would have been $\Sigma_{\rm gas} \sim 2\;\mpc$.
Results from the ATLAS$^{\rm 3D}$ survey also indicate that an initial gas surface density of this magnitude is plausible. 
\citet{Young2011} observed CO in early type galaxies. Their most gas rich galaxies had $\Sigma_{\rm gas} \sim 100\;\mpc$, but were found in low density environments.
More commonly, early type galaxies with detections and upper limits in their sample had $\Sigma_{\rm gas} \lesssim 10\;\mpc$. 
Moreover, for the galaxies in the ATLAS$^{\rm 3D}$ survey most similar to NGC~1272, the molecular gas dominated the total gas content \citep{Crocker2012}.

Thermal coronae of galaxies in clusters also indicate that the ISM is stripped to radii comparable to what we infer from the bent jets in NGC~1272. 
\citet{Sun2005} used deep Chandra observations to characterize hot gas bound to NGC~1265, which is one of the original bent-doubles.
They found that its corona extended only 0.8~kpc from the galaxy center in the direction of motion of NGC~1265, indicating a remaining gas reservoir even smaller in radial extent than what we find for NGC~1272.
Moreover, similar thermal x-ray coronae are found in a number of members of galaxy clusters \citep{Sun2007}.

If NGC~1272 initially had a gas density comparable to other elliptical galaxies, ram pressure could have removed material down to radii $\sim$2~kpc.
Ram pressure alone cannot, however, remove a cosmic gas fraction from deep within the galaxy.
Instead, some other process (e.g., AGN feedback in the galaxy) must initially remove a large fraction of gas in the interior of the galaxy before ram pressure can become effective.

\section{Conclusions}
NGC~1272, a massive elliptical galaxy near the center of the Perseus cluster, has an AGN powering a radio jet. 
The jet is bent back by ram pressure from the Perseus ICM, with a radius of curvature of $\sim$2~kpc.
The standard assumptions used in interpreting observations of synchrotron emission from bent-doubles appear to be inconsistent with the data for NGC~1272. 
If pressure support in the jets is provided by a roughly equal mix of relativistic particles and magnetic fields, the relativistic proton-to-electron energy ratio must be $k \sim 800$.
If this result is generally applicable, it is significant for attempts to measure gas densities via bent-doubles.
In this case, assuming a value $k = 1$ leads to an estimate of the gas density in the Perseus cluster that is too low by more than an order of magnitude. 

For the jets to feel ram pressure from the ICM, there must be minimal material in the ISM of NGC~1272 at radii $\sim$2~kpc and beyond.
We have shown that the same ram pressure that bends the jets could plausibly remove the ISM of the galaxy down to 2~kpc.
NGC~1272 thus provides a dramatic example of how a galaxy cluster's gas content can influence the evolution of its member galaxies. 

\section{Acknowledgments}
We thank the anonymous referee for constructive comments. We are grateful to Eliot Quataert and Carl Heiles for reading the manuscript and providing helpful comments and suggestions, and to Alexander Tchekhovskoy, Roger Blandford, Matt George, Steve Croft, Charles Hull, and Leo Blitz for useful conversations. 
J. M. received support from a National Science Foundation Graduate Research Fellowship. 
M. M. received support from the Thomas and Alison Schneider Chair in Physics at UC Berkeley.
This research used NASA's Astrophysics Data System Bibliographic Services, the SIMBAD database and the VizieR catalogue access tool, the NASA/IPAC Extragalactic Database (NED), and APLpy, an open-source plotting package for Python.
The National Radio Astronomy Observatory is a facility of the National Science Foundation operated under cooperative agreement by Associated Universities, Inc.

\bibliographystyle{mn2ealt}
\bibliography{ms}

\end{document}